# Dual-band coupling between nanoscale polaritons and vibrational and electronic excitations in molecules

A. Bylinkin[1,2], F. Calavalle[1], M. Barra-Burillo[1], R. V. Kirtaev[2], E. Nikulina[1], E. B. Modin[1], E. Janzen[3], J. H. Edgar[3], F. Casanova[1,10], L. E. Hueso[1,10], V. S. Volkov[4], P. Vavassori[1,10], I. Aharonovich[5,6], P. Alonso-Gonzalez[7,8], R. Hillenbrand*[9,10], and A. Y. Nikitin*[2,10]

1 CIC nanoGUNE BRTA, 20018 Donostia - San Sebastian, Spain

2 Donostia International Physics Center (DIPC), 20018 Donostia-San Sebastián, Spain

3 Tim Taylor Department of Chemical Engineering, Kansas State University Manhattan, KS 66506, USA

4 XPANCEO, Bayan Business Center, DIP, 607-0406, Dubai, UAE

5 School of Mathematical and Physical Sciences, University of Technology Sydney, Ultimo, New South Wales 2007, Australia

6 ARC Centre of Excellence for Transformative Meta-Optical Systems, Faculty of Science, University of Technology Sydney, Ultimo, New South Wales 2007, Australia

7 Departamento de Fisica, Universidad de Oviedo, 33006 Oviedo, Spain

8 Nanomaterials and Nanotechnology Research Center (CINN), 33940 El Entego, Spain

9 CIC nanoGUNE BRTA and EHU/UPV, 20018 Donostia-San Sebastián, Spain

10 IKERBASQUE, Basque Foundation for Science, 48009 Bilbao, Spain

*alexey@dipc.org, r.hillenbrand@nanogune.eu

Strong coupling (SC) between light and matter excitations such as excitons and molecular vibrations bear intriguing potential for controlling chemical reactivity, conductivity or photoluminescence. So far, SC has been typically achieved either between mid-infrared (mid-IR) light and molecular vibrations or between visible light and excitons. Achieving SC simultaneously in both frequency bands may open unexplored pathways for manipulating material properties. Here, we introduce a polaritonic nanoresonator (formed by h-BN layers placed on Al ribbons) hosting surface plasmon polaritons (SPPs) at visible frequencies and phonon polaritons (PhPs) at mid-IR frequencies, which simultaneously couple to excitons and atomic vibration in an adjacent molecular layer (CoPc). Employing near-field optical nanoscopy, we first demonstrate the co-localization of strongly confined near-fields at both visible and mid-IR frequencies. After covering the nanoresonator structure with a layer of CoPc molecules, we observe clear mode splittings in both frequency ranges by far-field transmission spectroscopy, unambiguously revealing simultaneous SPP-exciton and PhP-vibron coupling. Dual-band SC may be exploited for manipulating the coupling between excitons and molecular vibrations in future optoelectronics, nanophotonics, and quantum information applications.

# Main text

Strong coupling (SC) - light-matter interaction leading to the formation of new hybrid modes whose separation of energy levels is larger than the sum of their average linewidths - provides intriguing possibilities for controlling various material properties, such as e.g. carrier transport in organic semiconductors[1,2], magnetotransport in two-dimensional electron gases[3], or chemical reactivity changes in microcavities[4–7]. Conventional optical resonators, such as Fabry-Perot microcavities, are typically used to couple matter excitations with light. An approach to explore SC at nanoscale is to use polariton resonators, enabling SC between excitons and surface plasmon polaritons (SPPs) at visible frequencies[16,1], or between molecular vibrations and phonon polaritons (PhPs) at mid-infrared (mid-IR)[10,11] to THz frequencies[12,13]. Interestingly, the interaction between excitons and molecular vibrations plays an important role in the optical properties of organic semiconductors[14–16], particularly in the singlet fission. Achieving dual-band SC, i.e. SC between polaritons and both excitons and molecular vibrations simultaneously could provide new opportunities for controlling the state of matter at the nanoscale. However, excitonic and vibrational resonances emerge at significantly different energy domains separated by orders of magnitude, thus challenging the creation of strongly confined electromagnetic fields (hot spots) at visible and infrared frequencies - an essential ingredient for achieving SC at the nanoscale - at the same spatial position and with a similar size. Metallic antennas have already been used to enhance the light-matter interaction in both visible and mid-IR frequency bands[17], in particular, to combine surface-enhanced Raman and infrared spectroscopy[18,19]. However, although metallic antennas work relatively well in the visible range, it is desirable to find alternatives for achieving SC in the mid-IR frequency range due to their low quality factor at these frequencies.

Van der Waals (vdW) materials have recently emerged as a promising platform for exploring enhanced light-matter interactions at the nanoscale, as they support a large family of ultra-confined polaritons[24,2] from visible to THz frequencies. In addition, these materials can be engineered with nanoscale precision, allowing precise control of light-matter interactions at subwavelength scales[22]. Furthermore, the combination of metal slabs or antennas with vdW materials can lead to hybrid heterostructures that support electromagnetic modes at different frequency ranges[23] and potentially allow the creation of co-located visible and infrared hotspots to achieve dual-band SC.

Here we demonstrate by numerical simulations that SC can be achieved simultaneously between SPPs and electronic transitions at visible frequencies and between PhPs and molecular vibration at mid-IR frequencies. For an experimental study, we employed nanoresonators based on a heterostructures composed of metal (Al) ribbons and monoisotopic hexagonal boron nitride (h-BN) flakes, which support both SPPs at visible frequencies and PhPs at infrared frequencies. PhP resonances offer the advantage of being stronger and narrower as compared to SPP resonators[2], thus facilitating the achievement of nanoscale vibrational SC[10,11]. We verify the two hotspots and their co-localization by scattering-type scanning near-field optical microscopy (s-SNOM). By placing molecules on the nanoresonators within the corresponding hotspots and performing visible and mid-IR far-field spectroscopy, we provide direct experimental evidence for large mode splitting in both frequency ranges.

SC can be directly observed by visualizing the propagation of PhPs in a cavity-free (unpatterned) slab in contact with organic molecules[25,26]. To explore the possibility of observing dual-band SC, we thus first perform a theoretical study considering a cavity-free heterostructure formed by a 50 nm-thick layer of CoPc between a 50 nm-thick Al layer and a 75 nm-thick h-BN slab (Fig 1a). CoPc molecules support both a vibrational resonance at mid-IR frequencies (1525 cm$^{-1}$) and two excitonic resonances at visible frequencies (14400 cm$^{-1}$ and 16200 cm$^{-1}$) (Fig 1c,d). By placing the molecules between the two polaritonic materials (h-BN and Al) we expect to achieve highly confined mid-IR and visible electromagnetic fields inside the molecular layer (dashed lines in the left panel of Fig. 1a), thus guaranteeing a strong overlap of both fields and the molecules at the same spatial location. Since the plasma frequency of Al is $\omega_p = 10.83$ eV $\approx 8.7 \times 10^4$ cm$^{-1}$ [27,28], the Al slab supports SPPs below $\omega_p/\sqrt{2} = 5.4$ eV $\approx 1.4 \times 10^4$ cm$^{-1}$, i.e. in the whole visible range. The field of these SPPs is vertically confined on a length scale of $1/k_{SPP,z} \sim 100$ nm in the frequency range corresponding to the exciton resonances of CoPc. On the other hand, h-BN exhibits two mid-IR Reststrahlen bands - defined by the transverse (TO) and longitudinal (LO) optical phonons (785 to 845 cm$^{-1}$ and 1394 to 1650 cm$^{-1}$) - where PhPs are supported[29]. Interestingly, since the out-of-plane and in-plane dielectric permittivities of h-BN differ in sign in both RBs, the h-BN slab supports a set of PhPs modes that are typically denoted M0, M1, etc. (so-called hyperbolic polaritons)[30]. The fundamental M0 mode exhibits the longest wavelength and propagation length and thus is typically the dominating mode in h-BN nanoresonators[33,3]. In the chosen structure, the momentum of the M0 mode at the molecular-vibrational resonance of CoPc is comparable to that of SPPs in the Al layer at the exciton resonance of CoPc. As a result, the vertical field confinement of the fundamental M0-PhP mode outside of the slab ($1/k_{M0,z} \sim 80$ nm) is comparable to that of the SPPs, indicating that a molecular layer with a thickness of a few dozens of nm should be sufficient to achieve SC in both frequency bands.

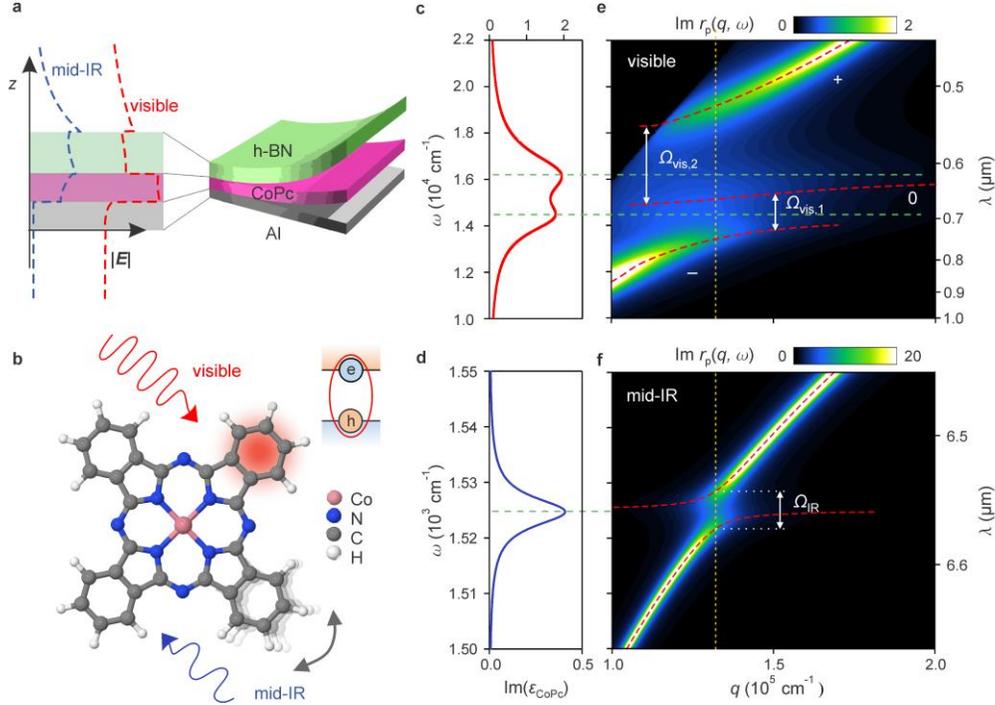

Figure 1. **Dual-band SC between polaritons and molecular excitations at both visible and mid-IR frequencies. a,** Schematic of the three-layer structure considered (50nm-Al/50nm-CoPc/75nm-h-BN). Left panel: blue and red dashed lines show the mode profile of the polaritonic modes in the mid-IR and visible frequency ranges, respectively (blue line for $\omega = 1530$ cm$^{-1}$, red line for $\omega = 19500$ cm$^{-1}$). **b,** Schematic of the atomic structure of the molecule CoPc together with both the vibrational and excitonic resonances in it. **c, d,** Experimentally extracted Im($\varepsilon_{CoPc}$) in the visible and mid-IR frequency ranges, respectively. **e, f,** Colour plots showing the calculated imaginary part of the Fresnel reflection coefficient of the structure in both frequency ranges. Red dashed lines show the calculated dispersion of the quasi-normal modes, assuming complex-valued frequency and real-valued momentum. Horizontal green dashed lines indicate the resonance frequencies of the molecular excitations.

To explore the concept of simultaneous dual-band SC, we calculated the dispersion of the polaritons in the cavity-free heterostructure, Fig 1a. To that end, we plot the imaginary part of the Fresnel reflection coefficient, Im[$r_p(q, \omega)$], at both visible and mid-IR frequencies (colour plots in Fig. 1e and 1f, respectively). The dashed red curves represent the dispersions of the quasi-normal polaritonic modes calculated from the poles of $r_p(q, \omega_c)$ in the plane of the complex-valued frequency $\omega_c = \omega - i\Gamma/2$, with $\Gamma$ being the mode linewidth (see Methods). Both colour plots and dispersion curves clearly reveal an anti-crossing (mode splitting) in both visible and mid-IR frequency ranges. To characterize the anti-crossing, we determine the mode splitting, $\Omega$, as the minimum vertical distance between the real part of the complex frequencies. The comparison of $\Omega$ with half the sum of the linewidths of the upper, $\Gamma_+$, and lower, $\Gamma_-$, coupled states determines whether the coupling between polaritons and molecular excitations is weak, $\Omega < (\Gamma_+ + \Gamma_-)/2$, or strong, $\Omega > (\Gamma_+ + \Gamma_-)/2$ (ref[33]). In the mid-IR frequency range, our calculation yields $\Omega_{IR} = 6.8$ cm$^{-1} \approx 8.4*10^{-4}$ eV and $\Gamma_{IR+,-} = 4$ cm$^{-1} \approx 5*10^{-4}$ eV, such that the SC condition is well fulfilled. In the visible range, the three initial excitations (SPP and two excitons) couple, leading to the formation of three hybrid states (red dashed lines in Fig.1e). By extracting the mode splitting as the minimum vertical distance between the real frequencies of the adjacent hybrid states, we obtain $\Omega_{vis,1} = 1.6*10^3$ cm$^{-1} \approx 0.2$ eV (between the low "–" and middle "0" branches) and $\Omega_{vis,2} = 3.9*10^3$ cm$^{-1} \approx 0.48$ eV (between the middle "0" and upper "+" branches). We found that the SC criterion is fulfilled for the split branches "0" and "+", since $\Omega_{vis,2} > (\Gamma_{vis,2,+} + \Gamma_{vis,2,0})/2$, where $\Gamma_{vis,2,+,0} \simeq 1.6*10^3$ cm$^{-1} \approx 0.2$ eV is the linewidth of the middle and upper hybrid states, respectively. On the other hand, the SC criterion is not fulfilled for the split branches "–" and "0". Altogether, our calculations reveal that the combination of different polaritonic materials in a single heterostructure can be used to achieve dual-band SC.

To demonstrate experimentally the dual-band SC between polaritons and CoPc molecular excitations, we first have to consider the need to access the large intrinsic momenta of polaritons, which in the case of far-field illumination requires the presence of leaky modes in the heterostructure. The latter can arise in open Fabry-Perot (FP) nanoresonators, which can be fabricated by (i) nanostructuring the polaritonic slab, e.g. in the form of ribbons[1,3], or (ii) nanostructuring the substrate below the h-BN slab[35], i.e. by refractive index engineering. Importantly, the latter option allows the use of a pristine h-BN slab, thus preserving its crystal quality. Interestingly, by placing a 75 nm-thick h-BN flake on an Al grating (Fig 2a), we can engineer FP nanoresonators by simultaneously applying concepts (i) and (ii). Indeed, while SPPs propagating across the Al ribbons (along the $x$-axis) in the visible range reflect directly at the Al edges (concept i), PhPs in the h-BN slab are reflected between the h-BN/air and h-BN/Al boundaries due to a refractive index step at mid-IR frequencies (concept ii). Fig. 2b,c show the normalised measured far-field extinction spectra of the nanoresonators in both frequency ranges. We can clearly recognize asymmetric peaks around $\omega = 1510$ cm$^{-1}$ and $\omega = 1.6*10^4$ cm$^{-1}$, which are identified (see Supplementary information SIII) as Fano-type FP resonances and, partially, as Bragg resonances arising from the overlap between the electromagnetic fields of adjacent cavities. As the inverse width of the Al ribbons ($w^{-1}$) corresponds to an effective momentum of the resonating polaritons, we fabricated a set of structures with different $w$, in order to cover a wide range of momenta (Supplementary information SII).

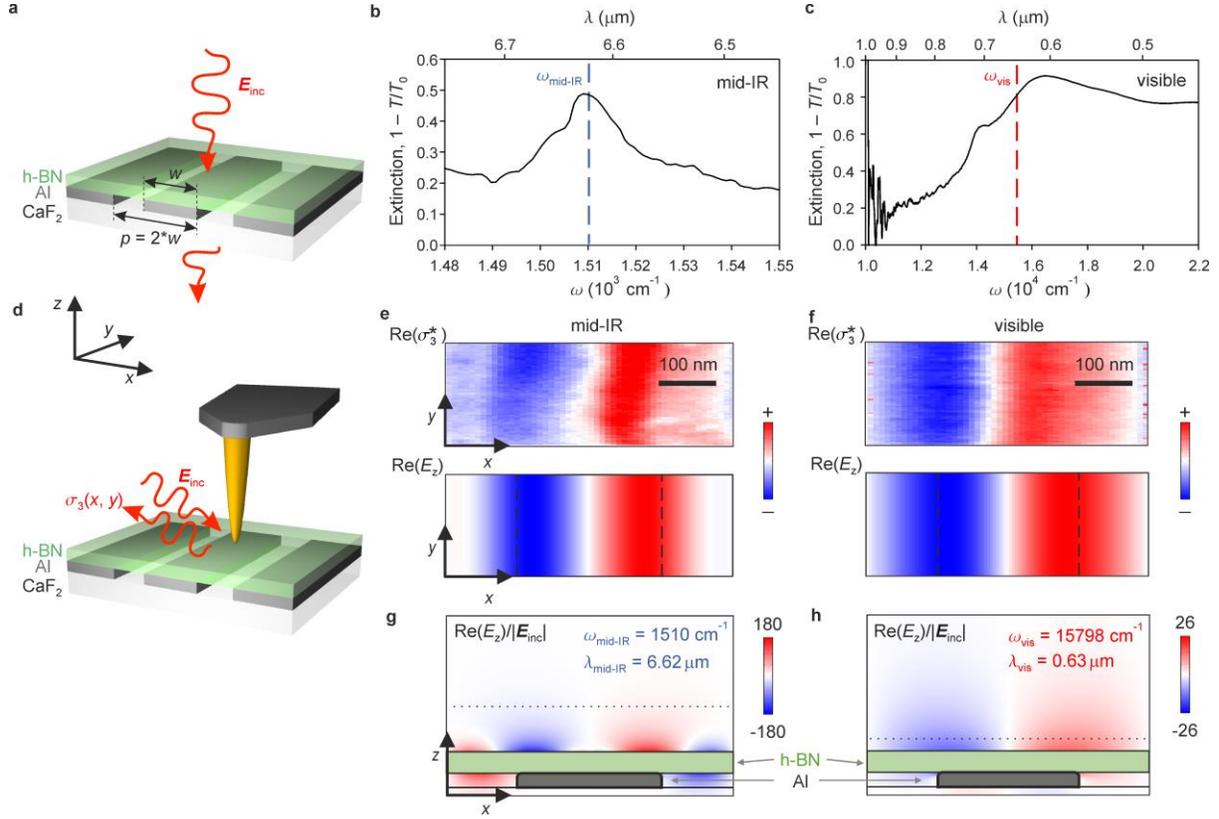

Figure 2. **Experimental demonstration of dual-band polaritonic nanoresonators. a,d** Schematics of the far-field experiments and s-SNOM optical characterization of the polaritonic nanoresonators. **b, c,** Experimental extinction spectrum of the nanoresonators created by employing an Al grating with period $p = 500$ nm in the mid-IR and visible frequency ranges, respectively. **e, f,** (top panels) Near-field images of the nanoresonators in the mid-IR and visible frequency ranges, measured at $\omega_{mid\text{-}IR} = 1510$ cm$^{-1}$ and $\omega_{vis} = 15798$ cm$^{-1}$ as indicated by the blue and red dashed lines in b,c, respectively. A clear dipolar fundamental mode is observed in both cases. (Bottom panels) Simulated Re($E_z$) in the $x$-$y$ plane, extracted 80 and 20 nm above the structure as indicated by the vertical dashed lines in g and h, respectively. Vertical dashed lines indicate the edges of Al ribbon. **g, h,** Simulated Re($E_z$)/|$E_i$| in the $x$-$z$ plane at the same frequencies. The period of the Al grating, the width of the Al ribbon and the thickness of the h-BN slab employed in the bottom panels **e, f** and **g, h** are $p = 500$ nm, $w = 250$ nm, and $d_{h\text{-}BN} = 75$ nm, respectively.

To corroborate our far-field spectroscopy experiments and to better understand the distribution of the near-field amplitude at the frequencies of the extinction peaks in Fig.2b,c, we performed s-SNOM nanoimaging of the nanoresonator heterostructure (see schematics in Fig. 2g,h) at mid-IR and visible frequencies (see Methods). In these nanoimaging experiments, the nanoresonators were illuminated with s-polarised light, whose electric field is perpendicular to the metal ribbons. The top panels in Fig. 2e,f show the resulting s-SNOM images for a nanoresonator when recording tip-scattered p-polarized light, which yields the $z$-component of the real part of the electric field at mid-IR ($\omega_{mid\text{-}IR} = 1510$ cm$^{-1}$, $\lambda_{mid\text{-}IR} = 6.62$ μm) and visible ($\omega_{vis} = 15798$ cm$^{-1}$, $\lambda_{vis} = 0.632$ μm) frequencies (see Methods and Supplementary Information SIV). Interestingly, despite the large difference in the wavelength of the incident light, we observe two bright areas with opposite polarity above the Al ribbons for both frequencies, revealing the excitation of transverse Fabry-Perot modes. To support these observations, we performed numerical simulations considering a nanoresonator with the experimental parameters and illuminating conditions. As can be clearly seen in the lower panels of Fig. 2e,f, excellent agreement is obtained between the simulated and experimental near-field distributions. In the normalized distributions of the vertical component of the electric fields (Fig. 2g,h), we identify strongly enhanced field amplitudes compared to the incident electric field, $E_{inc}$, thus confirming the formation of the hot spots. Therefore, we can conclude that the peaks observed in the far-field extinction spectra correspond to transverse Fabry-Perot resonances of PhPs at mid-IR frequencies and SPPs at visible frequencies. More importantly, the s-SNOM images in combination with the simulations confirm that both mid-IR and visible hot spots are spatially co-localized on the metal ribbons, thus potentially allowing for dual-band light-matter coupling involving the same molecules.

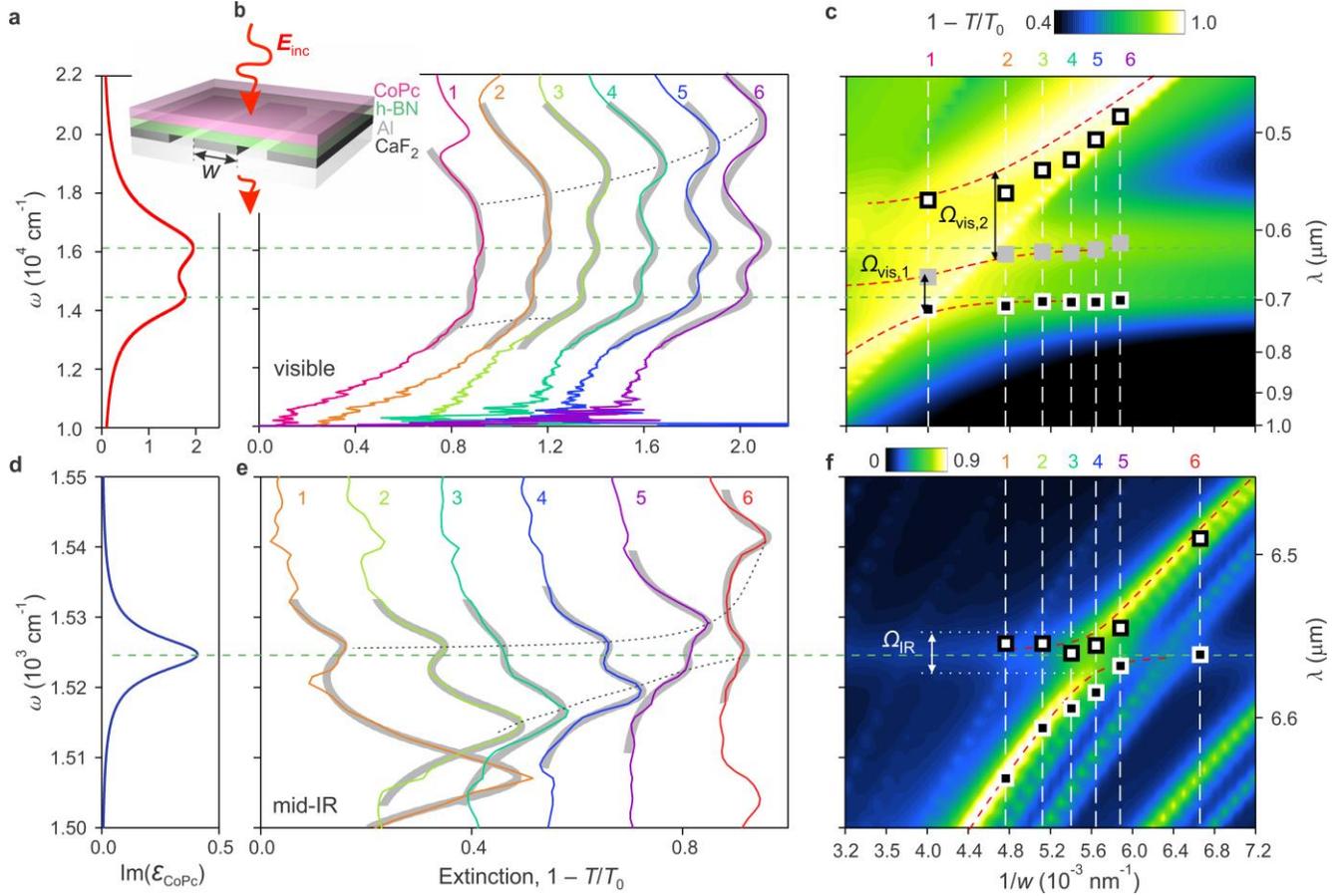

Figure 3. **Experimental demonstration of dual-band coupling between polaritons and vibrational and electronic molecular excitations. a, d,** Imaginary part of the permittivity of CoPc molecules ($Im(\varepsilon_{CoPc})$) experimentally extracted in the visible and mid-IR frequency ranges, respectively. **b, e,** Experimental extinction spectra of the polaritonic nanoresonators loaded with the molecular layer (see schematic) in the visible and infrared frequency ranges, respectively. Green dashed lines indicate the resonance frequencies of the molecular excitations. Black dashed lines are guides to the eyes indicating the resonance features in the extinction spectra corresponding to the coupled polaritonic modes. Thick grey lines in **b,e** represent fits using a three and two coupled harmonic oscillators model, respectively. **c, f,** Colour plots showing the simulated extinction of the nanoresonators in the visible and infrared frequency ranges, respectively. Red lines indicate the calculated dispersion assuming complex-valued frequency where $w^{-1} = 2q/(\pi - \alpha_{vis,IR})$ and $\alpha_{vis}$ = -0.12$\pi$, $\alpha_{IR}$ = -0.3$\pi$. The white and black squares in **f** represent the dispersion of the quasi-normal modes of the nanoresonators calculated using parameters obtained from the coupled oscillators fit in **e**.

To study experimentally the frequency splitting of the hybridized modes, we evaporated an 80 nm thick layer of CoPc molecules - for practical reasons - on Al/h-BN nanoresonators with different ribbon widths, $w$ (inset of Fig.

3), and measured the far-field extinction spectra in the visible and mid-IR frequency ranges (Fig 3b,e). We clearly observe peaks in the spectra whose positions depend on $w$, which can be attributed to the transverse politaritonic FP resonances that couple with the molecular excitations. More importantly, we observe an anti-crossing behaviour of the peaks (indicated by dashed black lines in Fig. 3b,e). These anti-crossings reveal coupling between SPPs and molecular excitons in the visible frequency range and coupling between PhPs and molecular vibrations in the mid-IR frequency range.

To corroborate the observed coupling between polaritons and molecular excitations, we simulated extinction spectra of nanoresonators with different ribbon widths (from 125 nm to 550 nm, colour plots in Fig. 3c,f) coated with a 80 nm thick CoPc layer. The colour plots in Fig. 3c,f show bright maxima, corresponding to polaritonic resonances, which shift to higher frequencies as $w^{-1}$ increases. We find good agreement between the trends in the frequency position of the maxima in the colour plot and the eye guides marking the maxima in the experimental extinction spectra (dashed black lines, Fig. 3b, e). In particular, in both the visible and mid-IR frequency ranges, the maxima in the simulated extinction spectra show anti-crossing behaviour around the frequencies of the molecular excitations (horizontal green dashed lines, Fig. 3), supporting our experimental observations.

To theoretically quantify the mode splitting, we performed a quasi-normal polaritonic mode analysis in a three-layer CoPc/Al/h-BN continuous heterostructure (see Methods). The calculated dispersions of the quasi-normal modes, $\omega(q)$, are shown on top of the simulated extinction spectra in Fig. 3c, f (dashed red lines). To compare the mode dispersion and the simulated extinction spectra, we related the real-valued in-plane polariton momenta, $q$, with the width of the grating, $w$, for each frequency range, $w^{-1} = 2q/(\pi \cdot \alpha_{vis,IR})$, where $\alpha_{vis}$= -0.12$\pi$, $\alpha_{IR}$ = -0.3$\pi$ are fit constants for the visible and mid-IR ranges, respectively. According to a simple FP model, the extracted values of the parameter $\alpha_{vis,IR}$ can be interpreted as the phase acquired by the polaritonic modes under reflection from the Al ribbon edges (SPP modes) and refraction index step in h-BN slab (PhP modes). We find excellent agreement between the calculated dispersions of the polaritons and the positions of the peaks in the extinction spectra in both spectral ranges. This agreement demonstrates that the interaction between polaritons and matter excitations in the continuous heterostructure is approximately equivalent to that in the nanoresonator heterostructure, which justifies that the analysis of a continuous heterostructure can be used to characterize the coupling parameters in the nanoresonator heterostructure. From the quasi-normal mode analysis we extract a mode splitting at mid-IR frequencies of $\Omega_{IR}$ = 5.9 cm$^{-1}$ $\approx$ 7.3*10$^{-4}$ eV, and a mode splitting at visible frequencies of $\Omega_{vis,1}$ = 1.1*10$^3$ cm$^{-1}$ $\approx$ 0.14 eV (between the lower and middle polariton branches) and $\Omega_{vis,2}$ = 3*10$^3$ cm$^{-1}$ $\approx$ 0.37 eV (between the middle and upper polariton branches). Considering the linewidths of the coupled states in the mid-IR frequency range, $\Gamma_{IR+,-}$ = 4 cm$^{-1}$ $\approx$ 5*10$^{-4}$ eV, and the linewidths in the visible frequency range, $\Gamma_{vis1,-,0}$ = 1.5*10$^3$ cm$^{-1}$ $\approx$ 0.19 eV and $\Gamma_{vis2,0,+}$ = 1.7*10$^3$ cm$^{-1}$ $\approx$ 0.21 eV, we find that the strong coupling criterion, $\Omega > (\Gamma_+ + \Gamma_-)/2$, is fulfilled both at mid-IR frequencies and for the "0" and "+" branches at visible frequencies.

To determine the dispersion of the quasi-normal modes and extract the value of the mode splitting from the experimental data, we fitted our extinction spectra in Fig. 3b,e using a classical model of coupled harmonic oscillators: two coupled oscillators in the mid-IR frequency range and three coupled oscillators in the visible frequency range, respectively (see Supplementary information SV). The coupled harmonic oscillator models allows us to reproduce the experimental extinction spectra by fitting the parameters of the uncoupled oscillators and the coupling strengths for each nanoresonator structure. With the parameters extracted from the fits we can calculate the dispersion of the quasi-normal modes of the coupled system, according to the dispersion equations in Supplementary information SV. In both spectral ranges, we find that the dispersion of the quasi-normal modes exhibits anti-crossing, as clearly shown by the squares in Fig.3cf. In the IR range, the minimum vertical distance between the dispersion branches of the modes yields a mode splitting of $\Omega_{IR}$ = 6.0 cm$^{-1}$ $\approx$ 7.3*10$^{-4}$ eV (Supplementary Information SV). In the visible spectral range, we find two mode splittings of $\Omega_{vis,1}$ = 1.1*10$^3$

cm$^{-1}$ ≈ 0.14 eV (between the lower and middle polariton branches) and $\Omega_{vis,2}$ = 2.1*10$^3$ cm$^{-1}$ ≈ 0.26 eV (between the middle and upper polariton branches). Thus, the extracted dispersions of the quasi-normal modes and the values of the mode splittings are in good agreement with the corresponding theoretical values for the continuous heterostructure in the mid-IR and visible spectral ranges (dashed red lines in Fig.3c,f) at the values of $w^{-1}$ corresponding to the fabricated structures (white vertical dashed lines in Fig. 3c,f). We note that the linewidths of the quasi-normal modes extracted from the fits in both frequency ranges are about 2-6 times larger than those calculated theoretically and vary for different structures, so that only the SC onset is reached. We explain this discrepancy by fabrication uncertainties and width variation along the Al ribbons, fabrication-induced roughness, defects, and the presence of higher order PhP modes in h-BN slab.

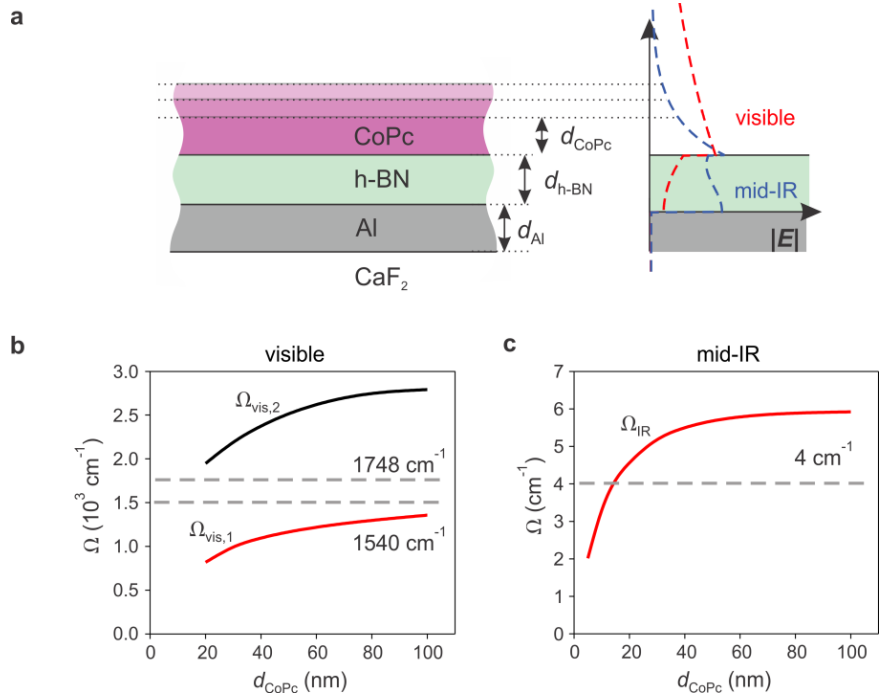

Figure 4. **Mode splitting dependence on molecular layer thickness. a.** Left: Schematic illustration of the Al/h-BN/CoPc heterostructure. Right: calculated electric field distribution in the heterostructure at $\omega$ = 1530 cm$^{-1}$ (blue dashed curve) and $\omega$ = 19500 cm$^{-1}$ (red dashed curve). **b,c,** Calculated mode splitting as a function of the molecular layer thickness in the visible and mid-IR frequency ranges. The square symbols in **c** are mode splitting extracted from the measured spectra. The Al and h-BN thicknesses are $d_{Al}$ = 50 nm, and $d_{h-BN}$ = 75. Horizontal dashed lines separate the regions of the strong and weak coupling. SC in both bands is observed for CoPc thicknesses larger than about 15 nm.

Finally, we analyze theoretically the coupling strength between polaritons and matter excitations. To do so, we extracted the mode splitting $\Omega$ for continuous heterostructures with different molecular layer thickness $d_{CoPc}$ from the previously developed quasi-normal analysis (see Methods). We find that the mode splitting, and thus the coupling strength between polaritons and molecular excitations, increases with $d_{CoPc}$ in both visible and mid-IR frequency ranges (Fig. 4b, c). This result is explained by a larger portion of the electromagnetic field of the polaritonic modes inside the molecular layer for larger thicknesses (Fig. 4a)[38,3]. Furthermore, we find that the mode splitting reaches saturation for a thickness of 60 nm, which is due to the full confinement of the PhP field inside the molecular layer (the field does not reach the air region). Interestingly, in the visible frequency range, the SC criterion is not fulfilled for the first exciton excitation even for a molecular layer thickness of 100 nm, while for the second excitonic excitation, the SC criterion is fulfilled for molecule layers thicker than 20 nm. In contrast, in the mid-IR range, the numerical calculations predict the observation of SC already for molecule layers as thin as 15 nm.

Our work demonstrates that engineering a heterostructure composed of plasmonic and phononic materials allows simultaneous access to light-matter interactions in the visible and mid-IR frequencies. Such heterostructure can be exploited to achieve dual-band strong light matter coupling, namely between nanoscale confined polaritons (SPPs and PhPs) and electronic or vibrational excitations of molecules. Momentum-energy coupling between excitations and polaritons can be achieved by tuning the dispersion of the latter through the thickness of the layers in the heterostructure. Future dual-band SC experiments may offer novel opportunities for manipulating chemical reactions, advanced optical imaging and sensing, or optomechanical up-conversion.

## Methods

### Sample preparation
Cobalt(II) Phthalocyanine, CoPc with sublimed quality (99.9%) (Sigma-Aldrich, Saint Louis, MO, USA) was thermally evaporated in an ultra-high-vacuum evaporator chamber (base pressure $<10^{-9}$ mbar) at a rate of 0.2 nm $s^{-1}$ using a Knudsen cell.

The h-BN crystal flake was grown from a metal flux at atmospheric pressure as described previously[38]. The thin layer used in this study was prepared by mechanical exfoliation with blue Nitto tape. Then, we performed a second exfoliation of the h-BN flakes from the tape onto a transparent polydimethylsiloxane (PDMS) stamp. Using optical inspection of the h-BN flake on the stamp, we identified a high-quality flake with appropriate thickness. This flake was transferred onto a $CaF_2$/Al gratings substrate using the deterministic dry transfer technique. To pick up and transfer this flake to another set of Al gratings we used a PDMS stamp with polycarbonate (PC) film, following the procedure described in ref [39].

Aluminium (Al) metal ribbon arrays (size of each array is 20 μm*20 μm) with different width of ribbons are fabricated using high-resolution electron beam (e-beam) lithography. The 50 nm of Al layer was e-beam evaporated onto the CaF2 substrate. Then negative resist (MA-N2401, 90 nm) was spin-coated followed by e-beam lithography of gratings (50 keV, 200 pA, dose $225 \div 375$ μC/cm$^2$) and resist development in AZ726 and reactive ion etching (RIE) of Al in $BCl_3$/$Cl_2$ plasma (pressure 40 mT, RIE power 100W). The resist was finally removed in $O_2$ plasma.

### Visible spectroscopy measurements

Transmission spectra in the visible (VIS) range were recorded using a wide-field optical microscope (Zeiss Axio). A broadband light source (400-1800 nm) is linearly polarised along the desired direction through a rotatable polarizer and used to illuminate the sample from the substrate in the Koehler configuration. After the polarizer, the light passes through an adjustable 4-blades slit and a condenser (NA 0.9) that projects the image of the slit on the sample surface, which is imaged by a CCD camera through a 50x polarisation-maintaining objective. By adjusting the slit blades, only the light transmitted by selected rectangular portions of the sample surface, as small as 20×20 μm$^2$, reaches the CCD detector for imaging. Once the desired area is selected, the transmitted light is diverted from the CCD and focused into a multicores optical fiber that convey the transmitted light to the VIS spectrometer (Ocean Optics USB2000+). VIS spectra are taken from the bare substrate, $T_0$, and from the area of interest, $T$, and the resulting normalised extinction spectra are obtained as 1-$T/T_0$.

**Fourier transform infrared spectroscopy measurements**

Infrared transmission spectra of the molecules, bare and molecule-coated heterostructure were recorded with a Bruker Hyperion 2000 infrared microscope (Bruker Optics GmbH, Ettlingen, Germany) coupled to a Bruker Vertex 70 FTIR spectrometer (Bruker Optics GmbH, Ettlingen, Germany). The normal-incidence infrared radiation from a thermal source (globar) was linearly polarised via a wire grid polarizer. The spectral resolution was 1 cm$^{-1}$.

**Eigenmode analysis**

We used the transfer matrix approach to calculate the quasi-normal modes[40]. They can be found by determining the poles in the Fresnel reflectivity of the layered sample for $p$-polarised light, $r_p$. To determine the poles, we numerically solved the equation 1/Abs($r_p$) = 0. We considered complex frequencies $\omega_c = \omega - i\Gamma/2$ and real-valued momenta $q$, and determined the poles of $r_p(q, \omega - i\Gamma/2)$, yielding $\omega(q)$, mode linewidth $\Gamma$. The dielectric permittivity of h-BN, CoPc, Al and CaF$_2$ were modelled as described in Supplementary Section I.

**Numerical simulation**

Full-wave numerical simulations using the finite-element method in the frequency domain (COMSOL) were performed to simulate the extinction spectra and study the electric field distribution around the heterostructure on top of a CaF$_2$ substrate. The dielectric permittivity of h-BN, CoPc, Al and CaF$_2$ were modelled as described in Supplementary Section I.

**Infrared and visible nanoimaging by s-SNOM**

We used a commercial s-SNOM set-up (Neaspec GmbH, Martinsried, Germany), in which the oscillating (at a frequency $\Omega_{tip} \cong 270$ kHz) metal-coated (Pt/Ir) atomic force microscope tip (Arrow-NCPt-50, Nanoworld, Nano-World AG, Neuchâtel, Switzerland) was illuminated by $s$-polarised mid-IR or visible radiations. We used tunable quantum cascade laser and He-Ne laser in the mid-IR and visible frequency ranges, respectively. The $p$-polarised backscattered light is recorded with a pseudoheterodyne Michelson interferometer. To suppress background scattering from the tip shaft and sample, the detector signal was demodulated at a frequency $3\Omega_{tip}$. We note that our imaging procedure (illuminating with $s$-polarised light and recording $p$-polarised light) allows substantially suppressing the excitation of polariton modes via the metallic tip.

# References

1. Orgiu, E. *et al.* Conductivity in organic semiconductors hybridized with the vacuum field. *Nat. Mater.* **14**, 1123–1129 (2015).

2. Hagenmüller, D., Schachenmayer, J., Schütz, S., Genes, C. & Pupillo, G. Cavity-Enhanced Transport of Charge. *Phys. Rev. Lett.* **119**, 1–6 (2017).

3. Paravicini-Bagliani, G. L. *et al.* Magneto-transport controlled by Landau polariton states. *Nat. Phys.* **15**, 186–190 (2019).

4. Thomas, A. *et al.* Tilting a ground-state reactivity landscape by vibrational strong coupling. *Science* **363**, 615–619 (2019).

5. Hutchison, J. A., Schwartz, T., Genet, C., Devaux, E. & Ebbesen, T. W. Modifying chemical landscapes by coupling to vacuum fields. *Angew. Chemie - Int. Ed.* **51**, 1592–1596 (2012).

6. Xiang, B. *et al.* Intermolecular vibrational energy transfer enabled by microcavity strong light–matter coupling. *Science* **368**, 665–667 (2020).

7. Yuen-Zhou, J., Xiong, W. & Shegai, T. Polariton chemistry: Molecules in cavities and plasmonic media. *J. Chem. Phys.* **156**, (2022).

8. Verre, R. *et al.* Transition metal dichalcogenide nanodisks as high-index dielectric Mie nanoresonators. *Nat. Nanotechnol.* **14**, 679–683 (2019).

9. Ji, W., Zhao, H., Yang, H. & Zhu, F. Effect of coupling between excitons and gold nanoparticle surface plasmons on emission behavior of phosphorescent organic light-emitting diodes. *Org. Electron.* **22**, 154–159 (2015).

10. Autore, M. *et al.* Boron nitride nanoresonators for Phonon-Enhanced molecular vibrational spectroscopy at the strong coupling limit. *Light Sci. Appl.* **7**, 17172–17178 (2018).

11. Autore, M. *et al.* Enhanced Light–Matter Interaction in 10B Monoisotopic Boron Nitride Infrared Nanoresonators. *Adv. Opt. Mater.* **9**, 1–9 (2021).

12. Damari, R. *et al.* Strong coupling of collective intermolecular vibrations in organic materials at terahertz frequencies. *Nat. Commun.* **10**, 1–8 (2019).

13. Kaeek, M., Damari, R., Roth, M., Fleischer, S. & Schwartz, T. Strong Coupling in a Self-Coupled Terahertz Photonic Crystal. *ACS Photonics* **8**, 1881–1888 (2021).

14. Bakulin, A. A. *et al.* Real-time observation of multiexcitonic states in ultrafast singlet fission using coherent 2D electronic spectroscopy. *Nat. Chem.* **8**, 16–23 (2016).

15. Musser, A. J. *et al.* Evidence for conical intersection dynamics mediating ultrafast singlet exciton fission. *Nat. Phys.* **11**, 352–357 (2015).

16. Aragó, J. & Troisi, A. Dynamics of the excitonic coupling in organic crystals. *Phys. Rev. Lett.* **114**, 1–5 (2015).

17. Domina, K. L., Khardikov, V. V., Goryashko, V. & Nikitin, A. Y. Bonding and Antibonding Modes in Metal–Dielectric–Metal Plasmonic Antennas for Dual-Band Applications. *Adv. Opt. Mater.* **8**, 1–6 (2020).

18. Le, F. *et al.* Metallic Nanoparticle Arrays : A Common. *ACS Nano* **2**, 707–718 (2008).

19. D'Andrea, C. *et al.* Optical nanoantennas for multiband surface-enhanced infrared and raman spectroscopy. *ACS Nano* **7**, 3522–3531 (2013).

20. Basov, D. N., Fogler, M. M. & García De Abajo, F. J. Polaritons in van der Waals materials. *Science* **354**, aag1992 (2016).

21. Low, T. *et al.* Polaritons in layered two-dimensional materials. *Nat. Mater.* **16**, 182–194 (2017).

22. Caldwell, J. D. *et al.* Photonics with hexagonal boron nitride. *Nat. Rev. Mater.* **4**, 552–567 (2019).

23. He, M. *et al.* Guided Mid-IR and Near-IR Light within a Hybrid Hyperbolic-Material/Silicon Waveguide Heterostructure. *Adv. Mater.* **33**, 1–9 (2021).

24. Caldwell, J. D. & Novoselov, K. S. Van der Waals heterostructures: Mid-infrared nanophotonics. *Nat. Mater.* **14**, 364–366 (2015).

25. Bylinkin, A. *et al.* Real-space observation of vibrational strong coupling between propagating phonon polaritons and organic molecules. *Nat. Photonics* **15**, 197–202 (2021).

26. Thomas, P. A., Menghrajani, K. S. & Barnes, W. L. Cavity-Free Ultrastrong Light-Matter Coupling. *J. Phys. Chem. Lett.* **12**, 6914–6918 (2021).

27. Rodrigo, S. G., García-Vidal, F. J. & Martín-Moreno, L. Influence of material properties on extraordinary optical transmission through hole arrays. *Phys. Rev. B - Condens. Matter Mater. Phys.* **77**, 1–8 (2008).

28. Rakić, A. D., Djurišić, A. B., Elazar, J. M. & Majewski, M. L. Optical properties of metallic films for vertical-cavity optoelectronic devices. *Appl. Opt.* **37**, 5271 (1998).

29. Geick, R., Perry, C. H. & Rupprecht, G. Normal modes in hexagonal boron nitride. *Phys. Rev.* **146**, 543–547 (1966).

30. Dai, S. *et al.* Tunable phonon polaritons in atomically thin van der Waals crystals of boron nitride. *Science* **343**, 1125–1129 (2014).

31. Alfaro-Mozaz, F. J. *et al.* Nanoimaging of resonating hyperbolic polaritons in linear boron nitride antennas. (2017) doi:10.1038/ncomms15624.

32. Caldwell, J. D. *et al.* Sub-diffractional volume-confined polaritons in the natural hyperbolic material hexagonal boron nitride. *Nat. Commun.* **5**, 1–9 (2014).

33. Törmö, P. & Barnes, W. L. Strong coupling between surface plasmon polaritons and emitters: A review. *Reports Prog. Phys.* **78**, 13901 (2015).

34. Ju, L. *et al.* Graphene plasmonics for tunable terahertz metamaterials. *Nat. Nanotechnol.* **6**, 630–634 (2011).

35. Lee, I. H., Yoo, D., Avouris, P., Low, T. & Oh, S. H. Graphene acoustic plasmon resonator for ultrasensitive infrared spectroscopy. *Nat. Nanotechnol.* **14**, 313–319 (2019).

36. Barra-Burillo, M. *et al.* Microcavity phonon polaritons from the weak to the ultrastrong phonon–photon coupling regime. *Nat. Commun.* **12**, 1–9 (2021).

37. Canales, A., Baranov, D. G., Antosiewicz, T. J. & Shegai, T. Abundance of cavity-free polaritonic states in resonant materials and nanostructures. *J. Chem. Phys.* **154**, (2021).

38. Liu, S. *et al.* Single Crystal Growth of Millimeter-Sized Monoisotopic Hexagonal Boron Nitride. *Chem. Mater.* **30**, 6222–6225 (2018).

39. Zomer, P. J., Guimarães, M. H. D., Brant, J. C., Tombros, N. & Van Wees, B. J. Fast pick up technique for high quality heterostructures of bilayer graphene and hexagonal boron nitride. *Appl. Phys. Lett.* **105**, (2014).

40. Passler, N. C. & Paarmann, A. Generalized $4 \times 4$ matrix formalism for light propagation in anisotropic stratified media: study of surface phonon polaritons in polar dielectric heterostructures. *J. Opt. Soc. Am. B* **34**, 2128 (2017).

# Acknowledgements

We acknowledge the Spanish Ministry of Science, Innovation and Universities (national projects PID2021-123949OB-I00, PID2021-122511OB-I00, PID2021-123943NB-I00, RTI2018-094861-B-I00, and the project CEX2020-001038-M of the Maria de Maeztu Units of Excellence Program), the Basque Government (grant numbers IT1164-19) and the European Union's Horizon 2020 research and innovation programme under the Graphene Flagship (grant agreement number 881603, GrapheneCore3) and the Doctoral Network "DYNAMO" (HORIZON-MSCA-2021-DN-01, project no. 101072818). Support for hBN crystal growth came from the Office of Naval Research, award number N00014-22-1-2582. I.A. acknowledges the Australian Research Council (CE200100010) and the Office of Naval Research Global (N62909-22-1-2028) for financial support. P.A.-G. acknowledges support from the European Research Council under starting grant no. 715496, 2DNANOPTICA, and the Asturias FICYT under grant AYUD/2021/51185 with the support of FEDER funds. A.Y.N., M. B.-B. and P.A.-G. acknowledge the Spanish Ministry of Science and Innovation (grants PID2020-115221GB-C42, MDM-2016-0618 and PID2019-111156GB-I00, respectively). This work is produced with the support of a 2022 Leonardo Grant for Researchers in Physics, BBVA Foundation. The Foundation takes no responsibility for the opinions, statements and contents of this project, which are entirely the responsibility of its authors.

# Author contributions

A.N. and I.A. conceived the study. The electron lithography of metal gratings was performed by R.K. supervised by V.S.V. Sample fabrication was performed by A.B., F. Calavelle and M.B.-B., supervised by F. Casanova and L.E.H. A.B. performed the experiments, data analysis and simulations. P.A.-G. contributed to the near-field imaging experiments. P.V contributed to the far-field transmission experiments in the visible. Electron microscopy images were performed by E.B.M. and E.A.N. E.J. and J.H.E. provided the isotopically enriched boron nitride. A.Y.N. and R.H. supervised the work. A.B. and A.Y.N. wrote the manuscript with the input of R.H., I.A. and P.A.-G. All authors contributed to scientific discussion and manuscript revisions.

# Competing interests

R.H. is co-founder of Neaspec GmbH, a company producing scattering-type scanning near-field optical microscope systems, such as the one used in this study. The remaining authors declare no competing interests.

Supplementary information for

"Dual-band coupling between nanoscale polaritons and
vibrational and electronic excitations in molecules"

## Table of Contents

# I.    Dielectric functions of materials

## A.    CoPc dielectric function in the infrared frequency range

We measured the relative infrared transmission spectrum of a 100 nm thick Cobalt(II) Phthalocyanine (CoPc) layer evaporated on top of a CaF$_2$ substrate. To extract the dielectric function of the CoPc molecules we calculate the relative transmission spectra, $T/T_0$, for the three layer system, using Fresnel coefficients [1], where $T$ is transmission through CaF$_2$/CoPc and $T_0$ is transmission through the CaF$_2$ substrate ($\varepsilon_{CaF_2} = 1.37$ in the considered infrared frequency range).

We modeled the dielectric function of the CoPc molecules by the Drude−Lorentz model assuming one classical harmonic oscillator to describe molecular vibrations in the considered infrared frequency range, as follows:

$$\varepsilon_{CoPc}(\omega) = \varepsilon_{\infty,IR} + \frac{S}{\omega_0^2 - \omega^2 - i\Gamma_{CoPc}\omega}, \qquad (S1)$$

where $\varepsilon_{\infty,inf}$ is a high-frequency dielectric constant, $S$ is a constant that is proportional to the effective strength of the Lorentz oscillator, $\omega_0$ and $\Gamma_{CoPc}$ represent the central frequency and the linewidth of the Lorentz oscillator, respectively. Fit yields $\varepsilon_{\infty,IR} = 2.8$ cm$^{-1}$, $S = 3600$ cm$^{-2}$, $\omega_0 = 1524.8$ cm$^{-1}$, $\Gamma_{CoPc} = 6$ cm$^{-1}$. With the parameters extracted from the fit, we are able to calculate the dielectric permittivity of CoPc molecules in the infrared frequency range, according to Equation S1, Figure S1.

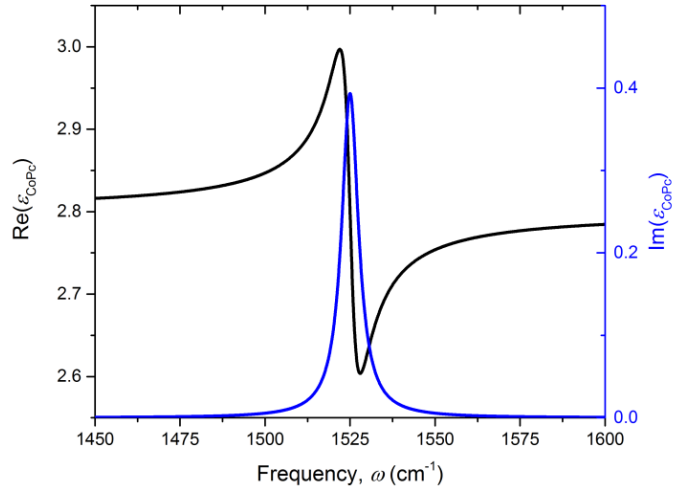

**Figure S1**. **The dielectric function of the CoPc molecules in the infrared frequency range**. Black and blue curves represent the real and imaginary parts of extracted CoPc dielectric function, respectively.

## B.    CoPc dielectric function in the visible frequency range

Following the same procedure as in the infrared frequency range, we measured the relative transmission spectrum for 20, 30, 50 and 100 nm thick Cobalt(II) Phthalocyanine (CoPc) layers evaporated on top of a CaF$_2$ substrate ($\varepsilon_{CaF_2} = 1.43$ in the considered visible frequency range). To extract the dielectric function of the CoPc molecules we used relative transmission spectra, $T/T_0$ calculated with the help of Fresnel coefficients.

We fit the dielectric function of the CoPc molecules by the Drude−Lorentz model assuming two classical harmonic oscillators to describe electronic transitions in visible frequency range, as follows:

$$\varepsilon_{CoPc}(\omega) = \varepsilon_{\infty,vis} + \frac{S_1}{\omega_{0,1}^2 - \omega^2 - i\Gamma_{CoPc,1}\omega} + \frac{S_2}{\omega_{0,2}^2 - \omega^2 - i\Gamma_{CoPc,2}\omega}, \qquad (S2)$$

where $\varepsilon_{\infty,vis}$ is a high-frequency dielectric constant, $S_k$  ($k = 1,2$) are constants proportional to the effective strength of the $k$th Lorentz oscillator, $\omega_{0,k}$ and $\Gamma_{CoPc,k}$ represent the central frequency and the linewidth of the $k$th Lorentz oscillator, respectively.

We fitted the relative transmission spectrum for each thickness of the molecular layer independently and then averaged the extracted fit parameters. The averaged fit parameters: $\varepsilon_{\infty,vis} = 1.6$ cm$^{-1}$, $\omega_{0,1} = 14402$ cm$^{-1}$, $S_1 = 31539456$ cm$^{-2}$, $\Gamma_{CoPc,1} = 1739$ cm$^{-1}$, $\omega_{0,2} = 16264$ cm$^{-1}$, $S_2 = 65755881$ cm$^{-2}$, $\Gamma_{CoPc,2} = 2369$ cm$^{-1}$. Figure S2 shows the dielectric permittivity of CoPc molecules in the visible frequency range calculated according to Equation S2 with the parameters extracted from the fit.

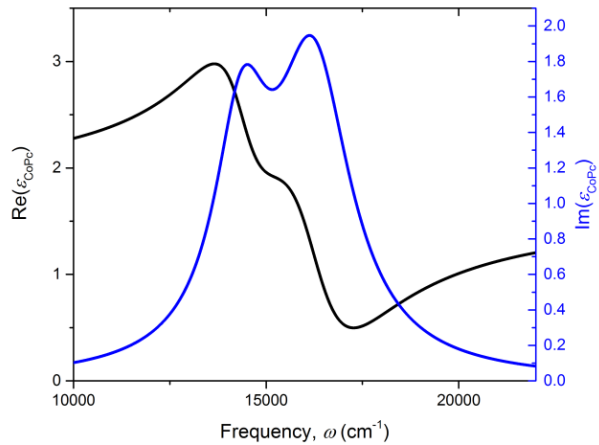

**Figure S2**. **The dielectric function of the CoPc molecules in visible frequency range**. Black and blue curves represent the real and imaginary parts of extracted CoPc dielectric function, respectively.

## C.    h-BN dielectric function

We used the isotopically ($^{10}$B) enriched h-BN [2]. The dielectric permittivity tensor of h-BN is modeled according to the following formula:

$$\varepsilon_{\text{h-BN},j}(\omega) = \varepsilon_{\infty,j} \left( \frac{\omega_{\text{LO},j}^2 - \omega^2 - i\omega\Gamma_j}{\omega_{\text{TO},j}^2 - \omega^2 - i\omega\Gamma_j} \right), \tag{S3}$$

where $j = \perp, \parallel$ indicates the component of the tensor parallel and perpendicular to the crystal axis, respectively. We took the parameters for the dielectric function of h-BN from ref.[2], except of $\varepsilon_{\infty,\perp}$. For the best matching of our near-field and far-field experiments we took $\varepsilon_{\infty,\perp} = 4$ instead of $\varepsilon_{\infty,\perp} = 5.1$ used in ref.3. We attribute this discrepancy to uncertainties introduced by the fabrication. All parameters for the dielectric function, which were used in the simulation, are presented in the Table S1.

| $j$ | $\varepsilon_\infty$ | $\omega_{\text{TO}}$, cm$^{-1}$ | $\omega_{\text{LO}}$, cm$^{-1}$ | $\Gamma$, cm$^{-1}$ |
|-----|-----|-----|-----|-----|
| $\perp$ | 4 | 1394.5 | 1650 | 1.8 |
| $\parallel$ | 2.5 | 785 | 845 | 1 |

Table S1. Parameters for the dielectric function of h-BN.

## D.    Al dielectric function

We modelled the Al dielectric function as a sum of Drude and Lorentz terms using parameters from the ref.[3].

## II.    Parameters of the arrays of Al ribbons

We fabricated the arrays of Al ribbons with different widths of the ribbons, $w$, and the periods of the structures, $p$, the latter designed to be twice the width of the ribbons, $p=2 \cdot w$. Figure S3 shows scanning electron microscope (SEM) images of the fabricated ribbon arrays. We extracted $p$ and $w$ from the measured SEM images and made sure that the filling factors, $f=w/p$, were indeed approximately 1/2 for all the fabricated structures. It is important to note that in Figure 3c,f of the main text and Figure S7, S8 of the supplementary information, to plot the quasi-normal modes of nanoresonator heterostructures we used doubled inverse period, $2 \cdot p^{-1}$, instead of the inverse ribbon width, $w^{-1}$. The extracted $p$, $w$ and calculated $f$, $p^{-1}$ parameters of the ribbon arrays are presented in Table S2

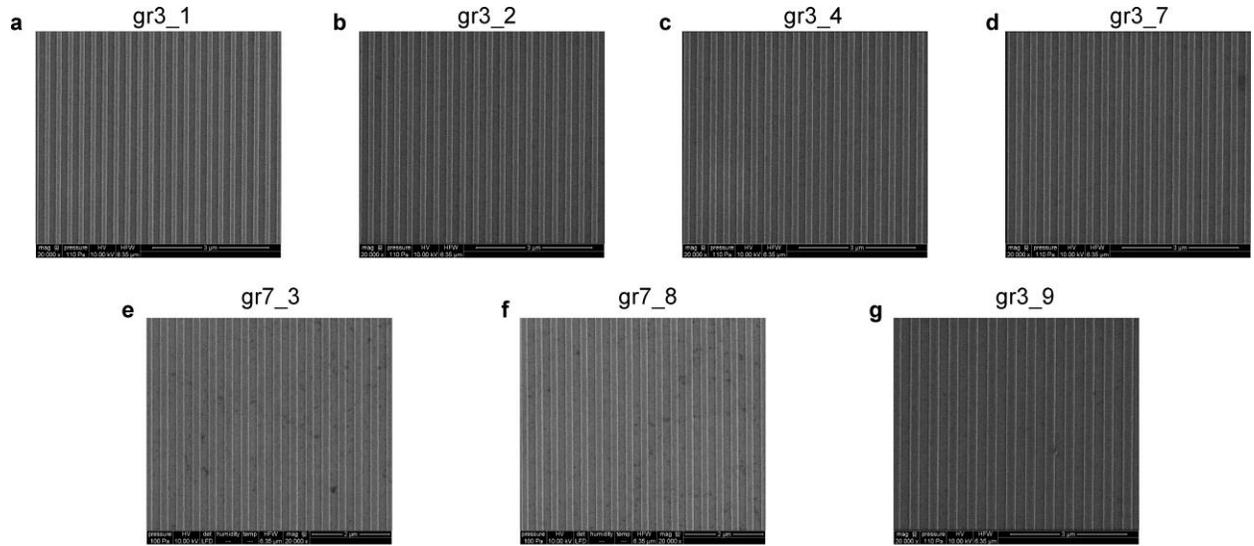

**Figure S3**. **SEM images of the fabricated arrays of Al ribbons, with the corresponding names of the arrays above the images.**

| Name of the array | $p$, nm | $w$, nm | $f = w/p$ | $1/p$, $10^{-3}$ nm$^{-1}$ |
|---|---|---|---|---|
| gr3_1 | 300 | 136 | 0.45 | 3.33 |
| gr3_2 | 339 | 160 | 0.47 | 2.94 |
| gr3_4 | 355 | 180 | 0.51 | 2.82 |
| gr3_7 | 370 | 190 | 0.51 | 2.70 |
| gr7_3 | 388 | 190 | 0.52 | 2.58 |
| gr7_8 | 419 | 220 | 0.49 | 2.39 |
| gr3_9 | 500 | 261 | 0.52 | 2.00 |

Table S2. Parameters of the fabricated arrays of Al ribbons, which are shown in Figure S3.

# III. Analysis of the polaritonic modes in the nanoresonator heterostructure

To study polaritonic modes in the nanoresonator heterostructure we performed the full-wave numerical simulation of electromagnetic fields using the finite-element method in frequency domain (COMSOL). We assumed a two dimensional (2D) geometry, namely an infinite number of infinitely long Al ribbons below the h-BN slab. We simulated transmission, reflection and scattering of a plane monochromatic wave normally incident onto the periodic array of nanoresonators. Figure S4a shows the schematics of one period of the simulated structure.

The right panels of Figure S4b,c show the calculated far-field extinction spectra, $1 - T$, where the $T$ is the power transmission coefficient in the visible and mid-IR frequency ranges, respectively. The calculated extinction spectra in both frequency ranges reveal numerous peaks. In the visible frequency range, we can clearly recognize a sharp peak in the extinction spectrum around $\omega = 1.4*10^4$ cm$^{-1}$ followed by a broad peak around $\omega = 1.65*10^4$ cm$^{-1}$ (Figure S4b, right panel). Both peaks form the so-called Wood-Rayleigh anomaly. Namely, the sharp peak (Rayleigh point) corresponds to the zero value of the $z$-component of the wavevector of the 1$^{st}$ order diffracted wave in the CaF$_2$ substrate. It takes place directly at the boundary between frequency regions where the 1$^{st}$ order diffracted wave has evanescent and propagating character. In contrast, the second peak (Wood anomaly) represents the first-order SPP Bragg resonance, corresponding to the pole in the transmission and reflection coefficients, and partially the Fabry-Perot (FP) resonance appearing as a result of multiple reflection of SPP mode (along the $x$-axis) from the edges of Al ribbons. In the mid-IR frequency range, the extinction spectrum manifests multiple peaks (Figure S4c, right panel). We assume that the latter emerge due to FP resonances appearing as a result of multiple reflection of PhP waveguiding modes from their refractive index steps defined along the $x$-axis by Al ribbons.

To analyze and interpret these peaks we generate the color plots (left panels of Figure S4b,c), representing the $z$-component of the electric field above the h-BN slab as a function of the frequency, $\omega$, and coordinate, $x$. In the visible frequency range, we observe the periodic field pattern along both the frequency and coordinate axes (Figure S4b, left panel). We see two bright localized areas ("hot spots") of different polarity along the coordinate axis, $x$, which can be explained by the presence of the transverse FP mode in the considered frequency range. We note that this FP mode can also be recognized in the simulated field distribution, Re($E_z$)/|$\mathbf{E}_i$|, in the $x$-$z$ plane at $\omega_{vis} = 15798$ cm$^{-1}$ (Figure 2h of the main text). We speculate that the periodicity of the field pattern along the frequency axis arises due to the complex interference between the electromagnetic fields of the FP mode and the incident and reflected waves. In the mid-IR frequency range, we find two frequency regions around $\omega = 1460$ cm$^{-1}$ and $\omega = 1510$ cm$^{-1}$ with the bright localized areas of the different polarity along the coordinate axis. These bright areas arise due to the presence of transverse FP resonances of PhP waveguiding modes at corresponding frequencies. These FP resonances can be characterized as "bright" modes since they have a nonzero in-plane dipole moment which can couple with propagating waves in free-space. As a result, these resonances appear as peaks in the far-field extinction spectrum (see the right panel of Figure S4c). In the color plot shown in the left panel of Figure S4c, we see that at $\omega = 1460$ cm$^{-1}$, a large portion of the mode volume is localized above the air region. In contrast, at $\omega = 1510$ cm$^{-1}$ the electric field of the mode is mainly localized above the Al ribbon. These observations allow us to assign the resonance around $\omega = 1460$ cm$^{-1}$ to the FP resonance of the PhP mode in h-BN slab above the air region. In contrast, a multi-peak in the extinction spectrum around the $\omega = 1510$ cm$^{-1}$ can be attributed to the FP resonances of the PhP modes above the Al ribbons. We explain the multi-peak structure in the extinction spectrum of the resonance around $\omega = 1510$ cm$^{-1}$ by the presence of higher-order PhP waveguiding modes in the h-BN layer. We finally identify the multi-peak resonance around $\omega = 1510$ cm$^{-1}$ as FP resonances and, partially, as Bragg resonance arising from the overlap between the electromagnetic fields of adjacent air and Al regions.

Note that, in contrast to the simulation, the experimental extinction spectrum in Figure 2b of the main text shows only the single resonant peak around $\omega = 1510$ cm$^{-1}$. This discrepancy can be explained by the fabrication uncertainties of the Al ribbon width throughout the array and by quality of the Al edges that play a crucial role in the far-field excitation of the higher-order PhP waveguiding modes in the h-BN slab.

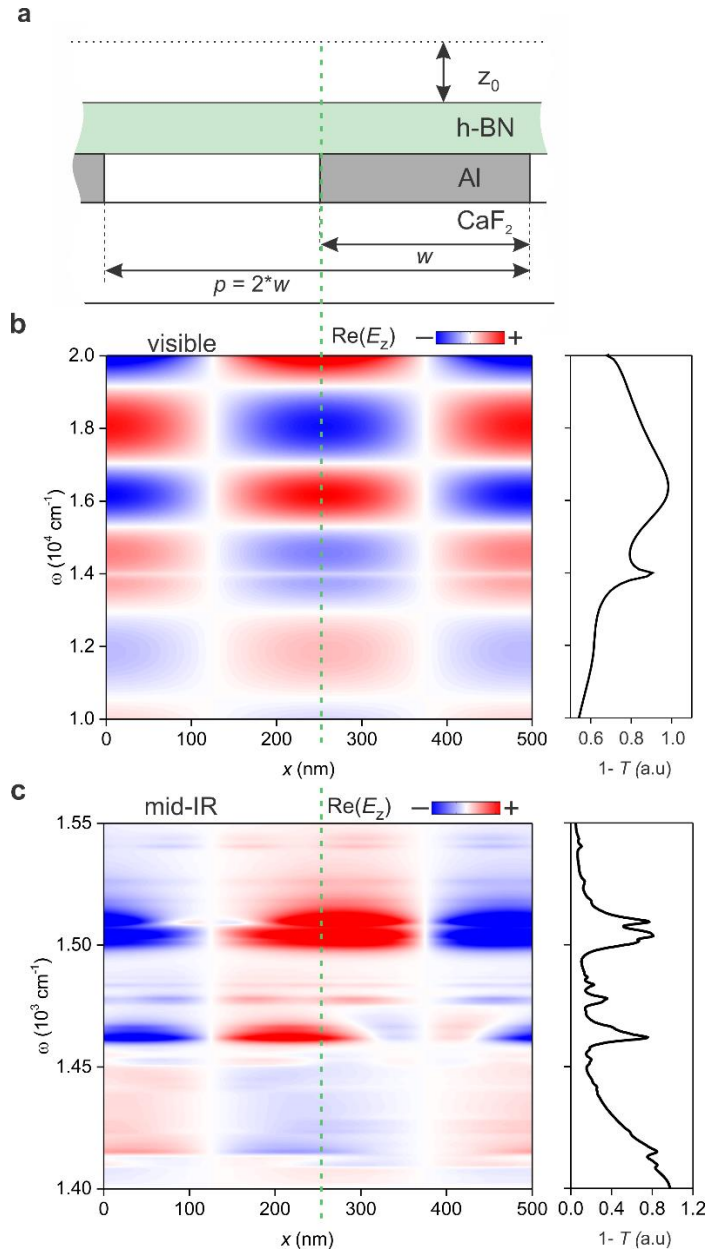

**Figure S4**. **Analysis of the polaritonic resonances in the visible and mid-IR frequency ranges**. **a.** Schematics of one period of the nanoresonator heterostructure with the period, $p$ = 500 nm, and Al ribbon width, $w$ = 250 nm. **b,c,** (left panel) Simulated electric field at the height of $z_0$ = 80 nm above the h-BN layer as a function of $\omega$ and $x$ coordinate in the visible and mid-IR frequency ranges, respectively. (right panel) Simulated extinction spectrum of the nanoresonator heterostructure in the visible and mid-IR frequency ranges, respectively.

# IV.  Data processing of infrared nanoimaging experiments

## A.  Mid-IR frequency range

Figure S5c,e,g show the raw amplitude, phase and real part (calculated using the amplitude and phase) of the complex-valued s-SNOM signal, $\sigma_3$. The data is represented as near-field images - the signal as a function of the tip position above the sample- of the set of 4 nanoresonators, which is schematically shown in Figure S5a,b. In order to reveal the mode field pattern above the nanoresonators, we subtracted the mean value of both the real and imaginary parts of signal for each horizontal line profile at the fixed $y$ coordinate. Figure S5d,f,g show the final data set of amplitude, phase and real part of the near-field images, $\sigma_3^*(x, y)$, after the subtraction of the mean values. The top panel in Figure 2e of the main text shows the final data of real part of the first resonator of Figure. S5h, for $x \in [0 : 0.5]\,\mu m$.

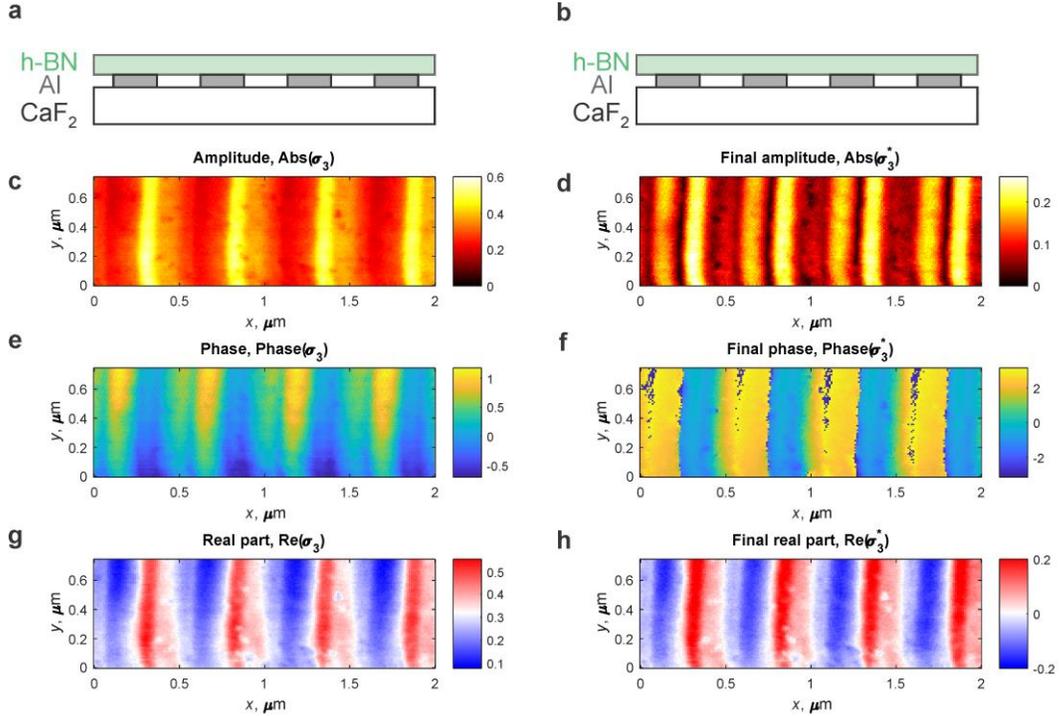

**Figure S5**. **Mean value subtraction from the near-field data in the mid-IR frequency range. a,b,** Schematics of the measured nanoresonator heterostructure. **c,e,g,** Raw amplitude, phase and real part of the near-field images of 4 nanoresonators, as indicated in the panels **a,b** in the mid-IR frequency range, measured at $\omega_{\text{mid-IR}} = 1510$ cm$^{-1}$, respectively. **d,f,h,** The resulting amplitude, phase and real part of the near-field images after the mean value subtraction, respectively.

## B.  Visible frequency range

Figure S6c,e,g show the raw amplitude, phase and real part (calculated using the amplitude and phase images) of the complex-valued s-SNOM signal, $\sigma_3$. Analogously to the mid-IR range, the near-field images are shown for the set of 4 nanoresonators, which is schematically shown in Figure S6a,b. We also analogously subtracted the mean value of both the real and imaginary parts of signal for each horizontal line profile at the fixed $y$ coordinate. Then we removed the propagating SPP Bloch mode by subtraction the complex signal $\sigma_B(x) = A_B e^{iGx + i\varphi_B}$ for each $y$ coordinate, where $A_B = 0.075$, $\varphi_B = 0.25\pi$ are real-valued fitting parameters and $G = \frac{2\pi}{p}$ is the Bragg vector of the array of Al ribbons with the period $p = 500$ nm. Figure S6d,f,g show the final data set of the amplitude, phase and real part of the near-field images, $\sigma_3^*(x, y)$, after the subtraction of the mean values and propagating SPP Bloch mode. The top panel in Figure 2f of the main text shows the final data of the real part of the first resonator of Figure S6h, for $x \in [0 : 0.5]\,\mu m$.

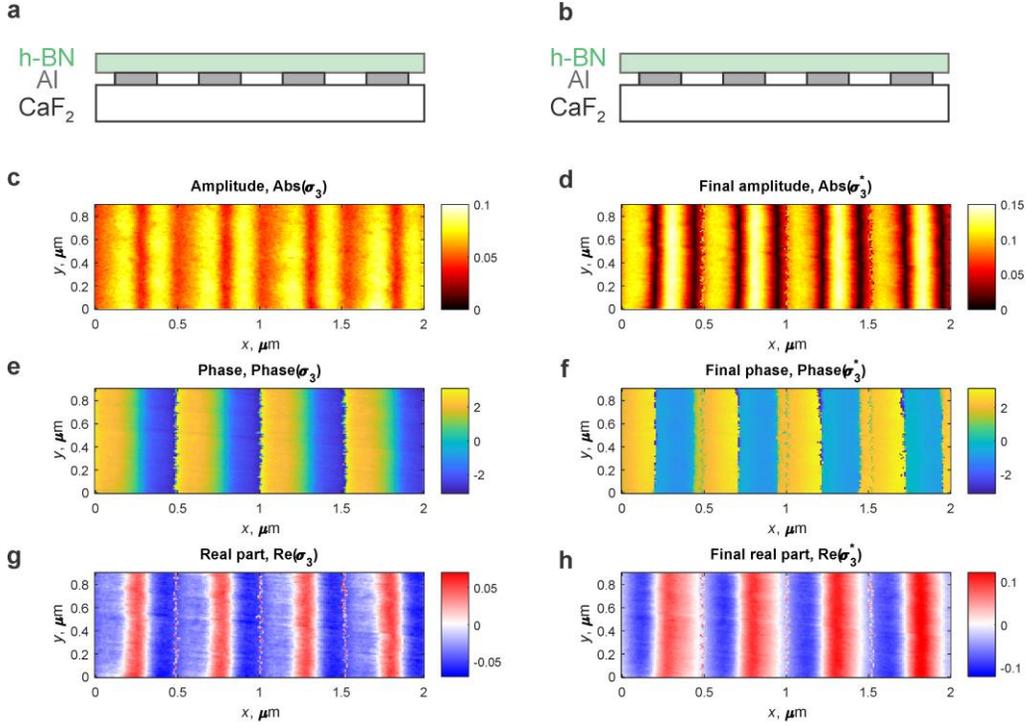

**Figure S6**. **Mean value and SPP Bloch mode subtraction from the near-field data in the visible frequency range. a,b,** Schematics of the measured nanoresonator heterostructure. **c,e,g,** Raw amplitude, phase and real part of the near-field images of 4 nanoresonators, as indicated in the panels **a,b** in the visible frequency range, measured at $\omega_{vis} = 15798$ cm$^{-1}$, respectively. **d,f,h,** The resulting amplitude, phase and real part of the near-field images, respectively.

# V. Coupled classical harmonic oscillators

## A. Mid-IR frequency range

In order to analyze the extinction spectra shown in Figure 3e of the main text, we phenomenologically described the coupling between the molecular vibration and the phonon polaritons via a model of classical coupled harmonic oscillators [4], [5]. The equations describing the motion of two coupled harmonic oscillators are given by[6]:

$$\begin{cases} \ddot{x}_1(t) + \Gamma_1 \dot{x}_1(t) + \omega_1^2 x_1(t) - 2g\dot{x}_p(t) = F_1(t) \\ \ddot{x}_{PhP}(t) + \Gamma_{PhP}\dot{x}_{PhP}(t) + \omega_{PhP}^2 x_{PhP}(t) + 2g\dot{x}_1(t) = F_{PhP}(t) \end{cases} \quad (S4)$$

where $x_{PhP}$, $\omega_{PhP}$ and $\Gamma_{PhP}$ represent the displacement, frequency and linewidth of the PhP mode, respectively. $x_1$, $\omega_1$ and $\Gamma_1$ represent the displacement, frequency and linewidth of the molecular vibration of CoPc, respectively. $F_{PhP}$ and $F_1$ represent the effective external forces that drive the motion of the oscillators. In the realistic electromagnetic problem the external electromagnetic field is the analog of the effective forces. $g$ represents the coupling strength between PhP mode and CoPc molecular vibration. From the oscillators model we can construct a quantity equivalent to the extinction, $C_{ext}$, of the analogous electromagnetic problem. It can be calculated as the work done by the external forces according to $C_{ext} \propto \langle F_{PhP}(t)\dot{x}_{PhP}(t) + F_1(t)\dot{x}_1(t)\rangle$ [4].

We fit $C_{ext}$ to the measaured extinction spectra of nanoresonator heterostructure with different ribbon width. Fits were performed according to $1 - \frac{T}{T_0} = C_{ext} + Offset$. In the fitting procedure, we take the same value of $\Gamma_1 = 6$ cm$^{-1}$ as we used in the CoPc dielectric function. $\omega_1$ was limited within a few wavenumbers from its initial value according to the CoPc dielectric function ($\omega_1 = \omega_{CoPc}=1524.8$ cm$^{-1}$), to allow for an eventual Lamb shift of the molecular vibration [7], [8]. $\omega_{PhP}$, $g$, $F_{1,PhP}$ and $\Gamma_{PhP}$ were considered as free parameters in all fits. The extracted values of the coupling strength for each fits are plotted as black symbols in the Figure S7b. With the parameters extracted from the fits we calculated the frequencies of the quasi-normal modes of the coupled system. The eigenfrequencies can be found from the dispersion relation, which arises from equaling the determinant of the system (S4) to zero [9] (assuming the harmonic time-dependence of the displacements, $e^{-i\omega t}$):

$$\omega_\pm + \frac{i\Gamma_\pm}{2} = \frac{\omega_{PhP} + \omega_1}{2} - i\frac{\Gamma_{PhP} + \Gamma_1}{4} \pm \frac{1}{2}\sqrt{4g^2 + \left(\omega_{PhP} - \omega_1 - i\frac{\Gamma_{PhP} - \Gamma_1}{2}\right)^2} \quad (S5)$$

We have used the approximation $\omega$-$\omega_j << \omega_j$, so that $\omega^2 - \omega_j^2 = 2\,\omega_j\,(\omega$-$\omega_j)$, with j=1, PhP. Figure S7a shows the calculated frequencies quasi-normal modes $\omega_\pm$ as a function of inverse width of nanoresonators, $1/w$. To find the mode splitting we first splined the real part of the calculated frequency of the quasi-normal modes (gray lines in Figure S7a) and then found the smallest vertical separation between the splined branches, $\Omega = 5.9$ cm$^{-1}$. All the fitting parameters are presented in Table S3.

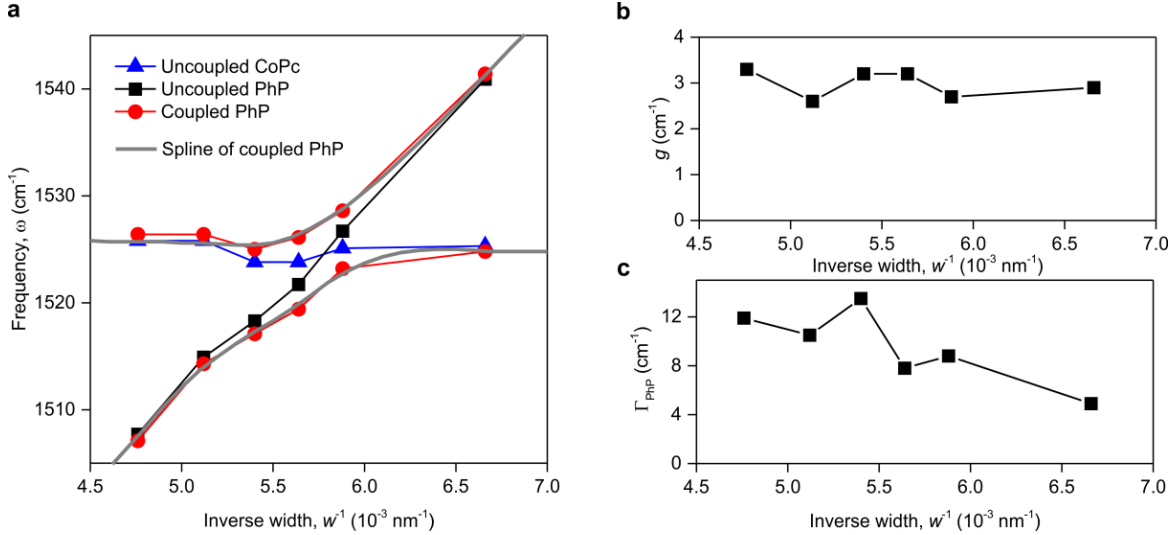

**Figure S7**. **Fitting the extinction spectra by the coupled oscillators model: mid-IR range. a,** Uncoupled frequencies of PhPs: $\omega_{\text{PhP}}$, and CoPc: $\omega_1$ (blue triangles and black squares, respectively), as extracted from the fits. Frequencies of the quasi-normal modes, $\omega_\pm$, calculated according to Eq. S6 (red circles). Grey solid lines show the spline of the eigenmodes frequencies as a function of inverse width of ribbons **b,** Black symbols show the coupling strength from the fits. **c,** The linewidth of uncoupled PhPs from the fits.

| Name of the array | $\omega_1$, cm$^{-1}$ | $F_1$, cm$^{-2}$ | $\omega_{\text{PhP}}$, cm$^{-1}$ | $\Gamma_{\text{PhP}}$, cm$^{-1}$ | $F_{\text{PhP}}$, cm$^{-2}$ | $g$, cm$^{-1}$ | *Offset* |
|---|---|---|---|---|---|---|---|
| gr7_8 | 1525.8 | 1.13 | 1507.7 | 11.9 | 2.16 | 3.3 | -0.38 |
| gr7_3 | 1525.8 | 1.21 | 1514.9 | 10.5 | 1.54 | 2.6 | -0.19 |
| gr3_7 | 1523.8 | 0.06 | 1518.3 | 13.5 | 1.82 | 3.2 | 0.07 |
| gr3_4 | 1523.8 | 0 | 1521.7 | 7.8 | 1.46 | 3.2 | 0.09 |
| gr3_2 | 1525.1 | 0 | 1526.7 | 8.8 | 1.37 | 2.7 | 0.1 |
| gr3_1 | 1525.3 | 0.4 | 1540.9 | 4.9 | 0.75 | 2.9 | 0.1 |

Table S3. Parameters of the coupled oscillators model to fit the experimental extinctions spectra in the mid-IR frequency range.

## B. Visible frequency range

In order to analyze the extinction spectra in the visible frequency range in Figure 3b of the main text, we phenomenologically described the coupling of the excitons and the surface plasmon polaritons (SPPs) via a classical model of three coupled harmonic oscillators, where we considered the coupling only between the SPP and two excitons. The equations of motion for the three coupled harmonic oscillators are given by:

$$\begin{cases} \ddot{x}_1(t) + \Gamma_1 \dot{x}_1(t) + \omega_1^2 x_1(t) - 2g_1 \dot{x}_{SPP}(t) = F_1(t) \\ \ddot{x}_2(t) + \Gamma_2 \dot{x}_2(t) + \omega_2^2 x_2(t) - 2g_2 \dot{x}_{SPP}(t) = F_2(t) \\ \ddot{x}_{SPP}(t) + \Gamma_{SPP} \dot{x}_{SPP}(t) + \omega_{SPP}^2 x_{SPP}(t) + 2g_1 \dot{x}_1(t) + 2g_2 \dot{x}_2(t) = F_{SPP}(t) \end{cases} \quad (S6)$$

where $x_{SPP}$, $\omega_{SPP}$ and $\Gamma_{SPP}$ represent the displacement, frequency and linewidth of the SPP mode, respectively. $x_{1,2}$, $\omega_{1,2}$ and $\Gamma_{1,2}$ represent the displacement, frequency and linewidth of the "1" and "2" excitons of CoPc, respectively. $F_{SPP}$ and $F_{1,2}$ represent the effective forces that drive their motions. $g_1$ represents the coupling strength between SPP mode and CoPc exciton "1", $g_2$ represents the coupling strength between SPP mode and CoPc exciton "2". Analogously to the mid-IR range, from the oscillators model we can construct a quantity equivalent to the extinction, $C_{ext}$, that can be calculated according to $C_{ext} \propto \langle F_{SPP}(t)\dot{x}_{SPP}(t) + F_1(t)\dot{x}_1(t) + F_2(t)\dot{x}_2(t)\rangle$.

We fit $C_{ext}$ to the measured extinction spectra of nanoresonator heterostructures with different ribbon width. Fits were performed according to $1 - \frac{T}{T_0} = C_{ext} + Offset$. In the fitting procedure, we take the same value of $\Gamma_{CoPc,1} = 1739$ cm⁻¹ and $\Gamma_{CoPc,2} = 2369$ cm⁻¹ as we used in the CoPc dielectric function. To minimize the number of fitting parameters in the visible frequency range, we fix the uncoupled frequencies of excitons $\omega_1 = \omega_{0,1} = 14402$ cm⁻¹ and $\omega_2 = \omega_{0,2} = 16264$ cm⁻¹ according to the CoPc dielectric function. $\omega_{SPP}$, $g_{1,2}$, $F_{1,2,SPP}$ and $\Gamma_{SPP}$ were considered as free parameters in all fits. The extracted values of the coupling strengths for each fits are plotted as black and red symbols in the Figure S8b.

With the parameters extracted from the fits we numerically calculated the three frequencies of the quasi-normal modes of the coupled system. The eigenfrequencies can be found from the dispersion relation, which arises from equaling the determinant of the system (S6) to zero (assuming the harmonic time-dependence of the displacements, $e^{-i\omega t}$). Figure S8a shows the three branches (red circles) of the real part of calculated frequencies of quasi-normal modes as a function of inverse width of the nanoresonators, $1/w$. To find the mode splitting, analogously to the mid-IR range, we first splined the real part of calculated frequencies of quasi-normal modes (gray lines in Figure S8a) and then extracted the smallest vertical separation between the adjacent splined branches $\Omega_{vis,1} = 1.1*10^3$ cm⁻¹ (between the lower and middle polariton branches) and $\Omega_{vis,2} = 3*10^3$ cm⁻¹ (between the middle and upper polariton branches). All the fitting parameters are presented in Table S4.

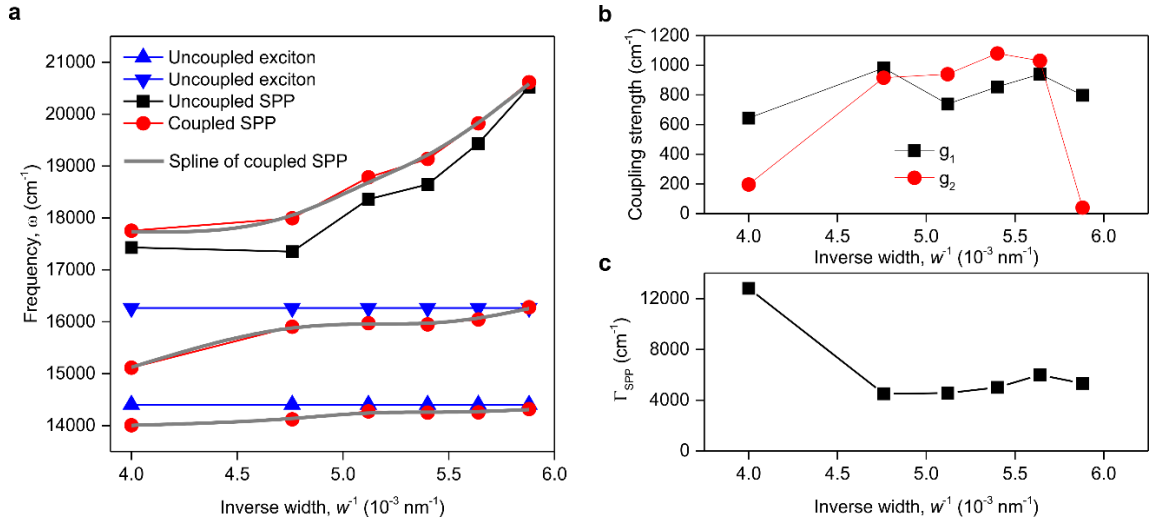

**Figure S8**. **Fitting the extinction spectra by the coupled oscillators model: visible range.** Uncoupled frequencies of SPPs, $\omega_{SPP}$, and CoPc excitons, $\omega_{1,2}$, (black squares and blue triangles, respectively) as extracted from the fits. Red circles show the eigenmode frequencies. Grey solid lines show the spline of the eigenmode frequencies as a function of inverse width of ribbons **b,** Black symbols show the coupling strength from the fits. **c,** The linewidth of uncoupled SPPs from the fits.

| Name of the array | $F_1$, cm$^{-2}$ | $F_2$, cm$^{-2}$ | $\omega_{SPP}$, cm$^{-1}$ | $\Gamma_{SPP}$, cm$^{-1}$ | $F_{SPP}$, cm$^{-1}$ | $g_1$, cm$^{-1}$ | $g_2$, cm$^{-1}$ | Offset |
|---|---|---|---|---|---|---|---|---|
| gr3_9 | 1600.65 | 100.18 | 17430 | 12801 | 15999.87 | 644.3 | 196.4 | -0.25 |
| gr7_8 | 1668.53 | 3000 | 17352 | 4515 | 5335 | 982 | 915.9 | 0.54 |
| gr7_3 | 1969.15 | 3073.72 | 18359.5 | 4552.7 | 6323.11 | 739.3 | 940 | 0.45 |
| gr3_7 | 2100.26 | 3504.32 | 18646.4 | 5006.3 | 7190.13 | 854.4 | 1080.3 | 0.35 |
| gr3_4 | 2397.8 | 3947.79 | 19433.8 | 5987.8 | 8429.8 | 939.9 | 1030.6 | 0.25 |
| gr3_2 | 2857.02 | 3996.99 | 20522.5 | 5317.9 | 7878.05 | 797.3 | 40 | 0.24 |

Table S4. Parameters of the coupled oscillators model to fit the experimental extinctions spectra in the visible frequency range.

# References

[1]     H. Raether, *Surface plasmons on smooth and rough surfaces and on gratings*. Springer, 2006.

[2]     A. J. Giles *et al.*, "Ultralow-loss polaritons in isotopically pure boron nitride," *Nat. Mater.*, vol. 17, no. 2, pp. 134–139, 2018, doi: 10.1038/NMAT5047.

[3]     S. G. Rodrigo, F. J. García-Vidal, and L. Martín-Moreno, "Influence of material properties on extraordinary optical transmission through hole arrays," *Phys. Rev. B - Condens. Matter Mater. Phys.*, vol. 77, no. 7, pp. 1–8, 2008, doi: 10.1103/PhysRevB.77.075401.

[4]     X. Wu, S. K. Gray, and M. Pelton, "Quantum-dot-induced transparency in a nanoscale plasmonic resonator," *Opt. Express*, vol. 18, no. 23, p. 23633, 2010, doi: 10.1364/oe.18.023633.

[5]     M. Autore *et al.*, "Boron nitride nanoresonators for Phonon-Enhanced molecular vibrational spectroscopy at the strong coupling limit," *Light Sci. Appl.*, vol. 7, no. 4, pp. 17172–17178, 2018, doi: 10.1038/lsa.2017.172.

[6]     L. Novotny, "Strong coupling, energy splitting, and level crossings: A classical perspective," *Am. J. Phys.*, vol. 78, no. 11, pp. 1199–1202, 2010, doi: 10.1119/1.3471177.

[7]     Y. Zhang *et al.*, "Sub-nanometre control of the coherent interaction between a single molecule and a plasmonic nanocavity," *Nat. Commun.*, vol. 8, no. May, pp. 1–7, 2017, doi: 10.1038/ncomms15225.

[8]     M. V. Rybin, S. F. Mingaleev, M. F. Limonov, and Y. S. Kivshar, "Purcell effect and Lamb shift as interference phenomena," *Sci. Rep.*, vol. 6, no. January, pp. 1–9, 2016, doi: 10.1038/srep20599.

[9]     K. Hennessy *et al.*, "Quantum nature of a strongly coupled single quantum dot-cavity system," *Nature*, vol. 445, no. 7130, pp. 896–899, 2007, doi: 10.1038/nature05586.